\documentclass[aps, pre,amsmath,amssymb,twocolumn,floatfix]{revtex4}

\newif\ifpdf\ifx\pdfoutput\undefined\pdffalse\else\pdfoutput=1\pdftrue\fi

\usepackage{graphicx}
\usepackage{bm}
\usepackage{epsfig}

\newcommand{\be}{\begin{equation}}

\newcommand{\ee}{\end{equation}}

\newcommand{\fex}{f^{\rm ex}}
\newcommand{\mi}{\rh_i} 
\newcommand{\sig}{\sigma} 
\newcommand{\rhsig}{\rh(\sig)} 
\newcommand{\intsig}{\int\!d\sig\,}
\newcommand{\rh}{\rho}  
\newcommand{\musig}{\mu(\sig)} 
\newcommand{\wi}{w_i(\sig)} 
\newcommand{\muex}{\mu^{\rm ex}} 
\newcommand{\pa}{^{(\alpha)}}
\newcommand{\rhzsig}{\rh^{0}(\sig)} 
\newcommand{\prior}{R(\sig)}
\newcommand{\fmom}{f_{\rm mom}}
\newcommand{\mze}{\rh_0}

\begin{document}

\title{\bf Liquid-gas coexistence and critical point shifts in size-disperse fluids}

\author{Nigel B. Wilding}
\affiliation{Department of Physics, University of Bath, Bath BA2 7AY, United Kingdom}

\author{Moreno Fasolo}
\author{Peter Sollich}
\affiliation{Department of Mathematics, King's College London, Strand, London WC2R 2LS, UK}

\date{\today}

\begin{abstract} 

Specialized Monte Carlo simulations and the moment free energy (MFE)
method are employed to study liquid-gas phase equilibria in
size-disperse fluids. The investigation is made subject to the
constraint of {\em fixed} polydispersity, i.e.\ the form of the
`parent' density distribution $\rho^0(\sigma)$ of the particle
diameters $\sigma$, is prescribed. This is the experimentally realistic
scenario for e.g.\ colloidal dispersions. The simulations are used to
obtain the cloud and shadow curve properties of a Lennard-Jones fluid
having diameters distributed according to a Schulz form with a large
($\delta\approx 40\%$) degree of polydispersity. Good qualitative
accord is found with the results from a MFE method study of a
corresponding van der Waals model that incorporates size-dispersity
both in the hard core reference and the attractive parts of the free
energy. The results show that polydispersity engenders considerable
broadening of the coexistence region between the cloud curves. The
principal effect of fractionation in this region is a common overall
scaling of the particle sizes and typical inter-particle distances, and
we discuss why this effect is rather specific to systems with Schulz
diameter distributions. Next, by studying a family of such systems
with distributions of various widths, we estimate the dependence of the
critical point parameters on $\delta$. In contrast to a previous
theoretical prediction,  size-dispersity is found to raise the critical
temperature above its monodisperse value. Unusually for a polydisperse
system, the critical point is found to lie at or very close to the
extremum of the coexistence region in all cases. We outline an argument
showing that such behaviour will occur whenever size
polydispersity affects only the range, rather than the strength of the
inter-particle interactions.


\end{abstract} 
\maketitle
\setcounter{totalnumber}{10}

\section{Introduction and background}

\label{sec:intro}

A fluid is termed polydisperse when its constituent particles are not
all identical, but exhibit continuous variation in some physical
attribute ($\sigma$, say) such as size, shape, charge etc. The state of
the system is then quantifiable in terms of a density distribution
$\rho(\sigma)$, measuring the number density of particles of each
$\sigma$ \cite{SALACUSE,KOFKE87}.  Conventionally, one identifies two
distinct classes of polydispersity: {\em variable} and {\em fixed}.
Variable polydispersity pertains to systems such as ionic micelles,
oil-water emulsions and blood
\cite{ISRAEL76,GRIMSON,DICKINSON,BARKER,BLAAK} where the degree of
polydispersity (as measured by the form of $\rho(\sigma)$) can change
under the influence of external factors such as pressure, temperature
or pH. By contrast, for systems exhibiting fixed polydispersity, the
relative proportions of particles of different $\sigma$ are prescribed by
the process of their chemical manufacture. Examples in this category
include colloidal dispersions, liquid crystals and polymer solutions
\cite{COLLOIDS,VROEGE93,NESARIKAR93}. For such systems, the {\em shape}
of $\rho(\sigma)$ is constant, only its {\em scale} can vary depending
on the quantity of solvent present.

In the present work we focus our attention on fluids exhibiting fixed
polydispersity. A central issue in such systems is the nature of their
phase behaviour. It has long been appreciated that this can be
considerably more complex than that of monodisperse systems
\cite{SOLLICH02}. Indeed, in some cases the presence of polydispersity
is predicted to engender completely new phases, such as the cascades of
demixing transitions that occur in length-disperse polymers
\cite{SOLLICH01} or the splitting of the triple point in polydisperse
liquid-crystal  polymers \cite{GHOSH03}. But even when
polydispersity is not a prerequisite for the existence of a phase
transition, (e.g.\ colloid-solvent demixing in colloids), it can
considerably enrich the character of the transition. The reason for
this is {\em fractionation}: at phase coexistence, particles of each
$\sigma$ may partition themselves unevenly between two (or more)
coexisting `daughter' phases  as long as--due to particle
conservation--the overall composition $\rho^0(\sigma)$ of the
`parent' phase is maintained. The consequences of fractionation for
phase diagrams can be dramatic.  For example, the conventional
liquid-gas binodal of a monodisperse system (which connects the ends of
tie-lines in a density-temperature diagram) splits into a `cloud' and a
`shadow' curve. These give, respectively, the density at which phase
coexistence first occurs and the density of the incipient phase. The
curves do not coincide because the shadow phase in general differs in
composition from the parent. Further fractionation effects are evident
when the phase diagram is represented in terms of external fields, such
as temperature and pressure. In contrast to monodisperse systems, for
which coexistence occurs along a line in the pressure-temperature
projection of the phase diagram, in the presence of polydispersity, this
line is broadened into a coexistence {\em region}
\cite{RASCON03,BELLIER00}.

Only recently has experimental work started to clarify in a systematic
fashion the generic consequences of polydispersity for phase
coexistence properties \cite{EVANS98,HEUKELUM03,FAIRHURST04}, and
fundamental questions remain unresolved. Consider, for instance, a
prototype model comprising a system of spherical particles exhibiting a
short ranged repulsion and a long ranged attraction. Even in such a
basic system, polydispersity may be manifest in different forms. The
simplest case is that of a purely size-disperse fluid in which the sole
variation amongst particles occurs in their diameters. Alternatively
the amplitude of interparticle interactions might vary between
particles.  More realistically, one should almost certainly consider a
coupled combination of both size and amplitude polydispersity i.e. the
strength with which a particle interacts with a neighbor depends on the
sizes of both. To date, however, there is relatively little known about
the individual effects of size and amplitude variations in systems
exhibiting fixed polydispersity in terms of their influence on phase
behaviour. For instance it is unclear just how the liquid-vapor
critical temperature and density are affected when polydispersity is
introduced into a model; how the critical point shifts depend on the
degree and functional form of the polydispersity; what is the extent
and nature of fractionation effects in the coexistence region and what
is the degree of associated coexistence curve broadening.  In view of
this, there clearly exists a need for systematic computational and
theoretical investigations of liquid-vapor phase behaviour in model
polydisperse fluids.  In the present paper, we address this need by
combining Monte Carlo (MC) simulation and analytical calculations to
investigate liquid-gas phase equilibria in size-disperse LJ fluids 

As regards previous studies of phase equilibria in polydisperse system,
we know of no prior simulation studies of fluids subject to fixed
polydispersity (although several have been performed for the variable
case \cite{STAPLETON,BOLHUIS,KOFKE93,KRISTOF}). Some analytical studies
do exist, however, and these typically seek to calculate the system
free energy as a function of a set of density variables. Unfortunately,
this task is complicated by the fact that the requisite free energy
$f[\rho(\sigma)]$ is a functional of $\rho(\sigma)$, and therefore
occupies an infinite dimensional space. Consequently the standard
tangent plane approach for identifying phase boundaries becomes
excessively unwieldy, both conceptually and numerically, and one is
normally obliged to resort to approximation
schemes~\cite{SOLLICH02}. For sufficiently simple model free energies
which generalize the van der Waals
(vdW) approximation to polydisperse systems, a direct attack on the
solution of the phase equilibrium conditions is nevertheless sometimes
possible, see e.g.~\cite{GUALTIERI82,XU00}. The reason for this is that
such models are normally ``truncatable'' so that the phase
equilibrium conditions can be reduced to nonlinear equations for a
finite number of variables (see sec.~\ref{sec:mfe} below). This approach
has been applied to the study of phase separation in
fluids exhibiting separate size and amplitude polydispersity, yielding
predictions for the cloud and shadow curves and critical parameters as
a function of polydispersity \cite{BELLIER00}. 

An approach which more systematically exploits the advantages of
truncatable models is the moment free energy method
\cite{SOLLICH01,SOLLICH02}. This approximates the full free energy
appropriately in terms of a ``moment free energy'' which depends on a
small number of density variables, thereby permitting the use of
standard tangent plane construction to locate phase boundaries. It thus
restores much of the standard geometrical insight to the phase
equilibrium analysis, while at the same time being computationally
efficient.  In particular, the method delivers (for the given free
energy) exact results for the location of critical points and the cloud
and shadow curves; other aspects of phase coexistence can be calculated
within a systematically refinable approximation scheme that eventually
allows one to locate the exact solution of the phase equilibrium
conditions. The moment free energy has been applied to the study of
phase behaviour in systems ranging from polydisperse hard rods to the
freezing of hard spheres \cite{SOLLICH02,FASOLO03}, and we shall deploy
it again in the present study.

The paper is arranged as follows~\cite{WILDING04}. We begin in
sec.~\ref{sec:formal} by  defining the form of the parent density
distribution that we have elected to study.  Next, in
sec.~\ref{sec:simmeth}, we introduce our simulation model and outline
the battery of specialized MC techniques necessary to cope with fixed
polydispersity and phase coexistence within a grand canonical ensemble
framework. The formalism underpinning the MFE method, together with the
form of the model free energy to which it has been applied, are
described in sec.~\ref{sec:mfe}.  Turning then to our results,
sec.~\ref{sec:results} compares the phase behaviour emerging from the
MFE calculations with that determined by the simulations. Finally,
sec.~\ref{sec:disc} presents a discussion of our findings and
highlights some issues meriting further investigation.

\section{Parent distribution}
\label{sec:formal}

We consider systems characterized by a single continuous polydisperse
attribute $\sigma$, with associated density distribution
$\rho(\sigma)$. Under conditions of fixed polydispersity, the {\em
shape} of $\rho(\sigma)$ is imposed. However, its {\em scale} can
vary depending on the amount of solvent present.  Following convention,
we define a ``parent'' distribution $\rho^0(\sigma)$ as

\begin{equation}
\rho^0(\sigma)=n_0f(\sigma)\:,
\end{equation}
where $n_0=N/V$ is the overall particle number density, while
$f(\sigma)$ is a nominated normalized shape distribution. Since
$\rho^0(\sigma)$ may vary only in terms of its scale $n_0$, the
system is constrained to traverse a so-called {\em dilution line} in the full
phase space of possible compositions. The value of $n_0$ thus serves to
parameterize this line.

The term ``parent'' is adopted to emphasize the fact that although (for a given
point on the dilution line) the density distribution is fixed globally,
the system might nevertheless phase separate into coexisting ``daughter"
phases whose density distributions differ from the parent.  The
daughter phase distributions are then related to the parent via a
simple volumetric average:

\begin{equation}
x^{(1)}\rho^{(1)}(\sigma) +x^{(2)}\rho^{(2)}(\sigma)=\rho^0(\sigma)
\label{eq:vav}
\end{equation}
where $Vx^{(1)}$ and $Vx^{(2)}$ are the volumes of phases $(1)$ and
$(2)$ respectively, with $V$ the sample volume. Note that on the cloud
curve marking the onset of phase coexistence, the volume fraction of
the incipient (shadow) phase is negligible and thus the density
distribution of the cloud phase coincides with that of the parent.

In this work, we will principally be concerned with unimodal shape distributions of the Schulz form:

\begin{equation}
f(\sigma)=\frac{1}{z!}\left(\frac{z+1}{\bar{\sigma}}\right)^{z+1}\sigma^z\exp\left[-\left(\frac{z+1}{\bar{\sigma}}\right)\sigma\right]\:.
\label{eq:schulz}
\end{equation}
Here $\bar\sigma\equiv 1$ is the average particle diameter, while $z$
is a parameter controlling the width of the distribution and hence
the dimensionless degree of polydispersity. The latter is conventionally defined as
the standard deviation of $\rho(\sigma)$, normalized by the mean:

\begin{equation}
\delta=\frac{\sqrt{\langle (\sigma-\bar\sigma)^2\rangle}}{\bar \sigma}\:.
\label{eq:delta}
\end{equation}

\section{Simulation aspects}

\label{sec:simmeth}

Possibly the simplest MC simulation strategy for obtaining the
thermodynamic properties of a polydisperse fluid is to populate a
simulation box of fixed volume $V$ with an appointed number of
particles $N$ whose sizes are drawn from the desired 
$\rho^0(\sigma)$ \cite{PHAN98}. Operationally, however, use of such
a canonical ensemble is far from optimal because it only samples a
single realization of the possible configurations of particle sizes.
Moreover the fixed overall particle number prevents effective study of
phase separation phenomena.  

Experience with the simulation of monodisperse fluids has shown that
use of the grand canonical ensemble (GCE) is highly effective for
studying fluid phase transitions \cite{WILDING95,PANAGIO00,WILDING01}.
Its application in the context of polydisperse fluids retains the
benefits of the monodisperse case. Moreover, it provides the key to
improved sampling of the density distribution $\rho(\sigma)$. This is
because $\rho(\sigma)$ can fluctuate {\em as a whole}, thus capturing
many different individual realization of the ensembles of particle sizes and
hence catering naturally for fractionation effects.  Notwithstanding
these advantages, however, the GCE might appear (at first sight)
unsuitable for the purpose of traversing a dilution line. This is because the
ensemble averaged density distribution $\bar{\rho}(\sigma)$ ostensibly
lies out-with the direct control of the simulator, its form being
instead determined by the imposed chemical potential distribution
$\mu(\sigma)$. Nevertheless, as we have shown previously
\cite{WILDING02}, it turns out to be possible to adapt $\mu(\sigma)$ in
such a way as to realize a specific desired form of $\rho(\sigma)$ for
any temperature of interest. By complementing this approach with
extended sampling MC techniques, phase coexistence properties can be
studied along a dilution line. The remainder of this section describes
the necessary techniques and the strategy for their implementation.

\subsection{Model and algorithm}

We consider a fluid of particles interacting via a pairwise
potential of the Lennard-Jones (LJ) form:
\begin{equation}
U(r_{ij}, \sigma_{ij})=4\epsilon\left[\left(\frac{\sigma_{ij}}{r_{ij}} \right)^{12}-\left(\frac{\sigma_{ij}}{r_{ij}} \right)^6\right]\:,
\label{Uij}
\end{equation}
with $\sigma_{ij}=(\sigma_i+\sigma_j)/2$. A cutoff was applied to the
potential for particle separations $r_{ij}>2.5\sigma_{ij}$.  The
polydispersity enters solely through the continuous distribution of diameters
$\sigma_i$, which were assigned the Schulz form Eq.~(\ref{eq:schulz}). 

The GCE Monte Carlo algorithm employed to study this model invokes four
types of operation: particle displacements, deletions, insertions, and
resizing. The particle diameter $\sigma$ is treated as a continuous
variable in the range $0\le \sigma \le \sigma_c$, with $\sigma_c$ a
cutoff which we choose
as $\sigma_c=3$ unless otherwise specified. However, 
distributions defined on $\sigma$ such as the instantaneous density $\rho(\sigma)$, and
the chemical potential $\mu(\sigma)$, are represented as histograms defined by
discretising the permitted range of $\sigma$ into $120$ bins. For
further details concerning the simulation algorithm, as well as the
structure, storage and acquisition of data, we refer the interested
reader to ref.~\cite{WILDING02}. 

The primary observables with which we shall be concerned are the
instantaneous density distribution $\rho(\sigma)$ and its ensemble
averaged form $\bar\rho(\sigma)$. We also accumulate the distribution of
several quantitites which do not depend explicitly on $\sigma$. These
are the fluctuating overall number density $p(n)$, where
$n=\int d\sigma \rho(\sigma)$, the distribution of the
fluctuating volume fraction $p(\eta)$, with $\eta=(\pi/6)\int d\sigma
\sigma^3\rho(\sigma)$, and the distributions of the configurational
energy $p(E)$ and the energy per particle $p(u)$ with $u=E/N$. 

For simulations of fixed polydispersity at some given temperature $T$,
one requires that form of the chemical potential distribution
$\mu(\sigma)$ for which $\bar\rho(\sigma)$ matches some target, namely
the prescribed parent $\rho^0(\sigma)$. The task of determining the
desired $\mu(\sigma)$ is complicated by the fact that it is an unknown
{\em functional} of the parent (i.e.\
$\mu(\sigma)=\mu[\rho^0(\sigma)]$). Recently, however, a computational
tool has been developed that efficiently overcomes this difficulty. The
non-equilibrium potential refinement (NEPR) scheme \cite{WILDING03}
enables the efficient iterative determination of
$\mu[\rho^0(\sigma),T]$, from a single simulation run, and without the
need for an initial guess of its form. To achieve this, $\mu(\sigma)$
is continually updated (in the course of a simulation run) in such a
way as to minimize the deviation of the instantaneous density
distribution $\rho(\sigma)$ from the desired parent form.  This
procedure realizes a non-equilibrium steady state for which
$\bar\rho(\sigma)=\rho^0(\sigma)$. However, since tuning $\mu(\sigma)$
`on-the-fly' in this manner violates detailed balance, the form of
$\mu(\sigma)$ thus obtained is not the equilibrium solution one
actually seeks. Nevertheless by performing a series of iterations in
which the degree of modification made to $\mu(\sigma)$ at each step is
successively reduced, one can drive the system towards equilibrium,
thereby obtaining the correct $\mu[\rho^0(\sigma),T]$. 

Notwithstanding the utility of the NEPR method for solving the inverse
problem $\mu[\rho^0(\sigma)]$, its deployment to determine the form of
$\mu(\sigma)$ all along a dilution line would represent a considerable
computational endeavor. Fortunately, however, this is not necessary. A
single application of the NEPR method, to obtain $\mu(\sigma)$ at a
state point on the dilution line close to the region of interest,
suffices to bootstrap an efficient tracking procedure based on
Histogram Extrapolation \cite{HR} (HE) techniques. Use of HE permits a
stepwise exploration of the dilution line to lower or higher densities.
The essential idea is to statistically reweight the data for
$\bar\rho(\sigma)$ measured at some dilution line point (given by the
value of $n_0$) in order to estimate the form of $\mu(\sigma)$
corresponding to some other not-too-distant value of $n_0=n_0^\prime$.
This is achievable, within the HE scheme, by minimizing a cost function
measuring the deviation of the extrapolated form of
$\bar\rho(\sigma)[\mu(\sigma)]$ from a target corresponding to the
desired parent $n_0^\prime f(\sigma)$. Full details of the tracking
procedure can be found in ref.~\cite{WILDING02}.

\subsection{Coexistence curve mapping strategy}

Simulation studies of phase coexistence present distinctive
computational challenges. Principal among these is the large free
energy (surface tension) barrier separating the coexisting phases. In
order to accurately locate coexistence points, a sampling scheme must
be utilized which enables this barrier to be surmounted \cite{BRUCE03}.
One such scheme is multicanonical preweighting \cite{BERG92}, which
utilizes a weight function in the MC acceptance probabilities, in order
to encourage the simulation to sample the interfacial configurations of
low probability. At a given coexistence state point, the requisite
weight function takes the form of an approximation to the inverse of
the distribution of the fluctuating number density, $p(n)$. While
specialized techniques allow determination of $p(n)$ from scratch, in
situations where one wishes to track a fluid-fluid phase boundary prior
determination of a weight function is unnecessary, provided one
commences from the vicinity of the critical point where the barrier to
inter-phase crossings is small \cite{WILDING01}. Data accumulated here
can be used (together with HE) to provide estimates of suitable
multicanonical weight functions at lower temperatures \cite{WILDING95}
where the barrier height is greater.

We seek the intersection of the dilution line with the temperature
dependent coexistence region. The latter is delineated by the cloud
curves which herald the onset of coexistence when approached from the
respective pure phases. The cloud curves (and their corresponding
shadow curves) were obtained as follows. The dilution line tracking
procedure was bootstrapped at a gas phase state point on the dilution
line at a moderately low temperature by using the NEPR method
\cite{WILDING03} to determine the form of $\mu(\sigma)$. Starting from
this point, the dilution line was then followed towards increasing
density (with the aid of HE) until the gas spontaneously liquefied.
Having estimated the location of a coexistence state point in this
manner, the temperature was increased in steps (whilst remaining on the
dilution line) until the density difference between the gas and the
spontaneously formed liquid vanished, signalling the proximity of the
critical point. Finite-size scaling methods \cite{WILDING95} were then
used to furnish more precise estimates for the critical parameters.

Having located the critical point, a detailed mapping of the cloud and
shadow curves was performed. The key to achieving this is the form
of the fluctuating overall number density, $p(n)$. The gas phase cloud
point (incipient liquid phase) corresponds to the situation where
$p(n)$ is bimodal, but with vanishingly small weight in the liquid
peak. Under these conditions, the position of the low density gas peak
provides an estimate of the gas phase cloud density, while that of the
liquid peak gives the liquid shadow density. The converse is true
for the liquid phase cloud point and its gas shadow. Determining the cloud
and shadow points as a function of temperature yields the cloud and
shadow curves. One tracks the gas and liquid cloud curves (and their
shadows) in a stepwise fashion downwards in temperature from the
critical point, using HE to negotiate each temperature step. HE yields
estimates not only for the form of $\mu(\sigma)$ on the cloud curve at
the next temperature, but also the requisite multicanonical weight
function. It should be pointed out that while the positions of the peaks
in $p(n)$ provide an accurate estimate of cloud and shadow points at
low temperatures, this breaks down near the critical point due to
finite-size effects \cite{WILDING95}. Thus a naive extrapolation of our
curves to their intersection point will tend to overestimate the
critical temperature. However, our independent determination of the
critical point using finite-size scaling methods is considerably more
accurate. 

In order to explore the coexistence region one must traverse the
dilution line connecting the cloud points at some desired subcritical
temperature. The strategy for doing so commences at one cloud point
and entails constructing a series of simulations in which the value of
$n_0$ is successively increased. The requisite multicanonical weight
function for each simulation state point is again found with the help of
HE from the previous state point.

\section{Moment free energy method}
\label{sec:mfe}

\subsection{Model}

To describe our system theoretically, we need a suitable free energy;
this is a functional of the density distribution $\rho(\sigma)$ and an
ordinary function of temperature. The free energy (density) of a
polydisperse system can generally be decomposed as
\begin{equation}
f = \int d\sigma \rho(\sigma) \left[\ln\rho(\sigma)-1\right] + \fex
\label{eq:fre_ene}
\end{equation}
where the first part is the free energy of an ideal polydisperse
mixture.  To find 
a suitable model the excess free energy $\fex$, we
approximate the repulsive part of our LJ interaction as completely
hard. For the corresponding contribution to $\fex$ we use the BMCSL
approximation $\fex_{\rm
BMCSL}$~\cite{BOUBLIK,MANSOORI,SALACUSE}. In the monodisperse
case this reproduces the Carnahan-Starling equation of
state~\cite{CarSta69}; in the general polydisperse case it is a
function of the moments up to third order of the density distribution,
$\rho_i = \int\! d\sigma\, \sigma^i\rho(\sigma)$ ($i=0\ldots 3$). We
then treat the {\em attractive} part of the interaction potential in the
simplest possible way, by adding a quadratic van der Waals term to
$\fex$. Using the fact that the interaction volume of two particles of
diameters $\sigma$ and $\sigma'$ scales as $(\sigma+\sigma')^3$, this
gives for the overall excess free energy
\begin{eqnarray}
\fex &=& \fex_{\rm BMCSL} - \frac{1}{2t} \int\! d\sigma\, d\sigma'\, 
\rho(\sigma) \rho(\sigma') (\sigma+\sigma')^3
\label{fex_first}
\\
&=& \fex_{\rm BMCSL} - \frac{1}{t}(\rho_0 \rho_3 + 3\rho_1 \rho_2)
\label{fex}
\end{eqnarray}
where $t$ is an appropriate dimensionless temperature. We note in this
context that for our theoretical calculations we measure all diameters
in units of $\bar\sigma$, all densities in units of the inverse volume
of a reference particle, $1/[(\pi/6)\bar\sigma^3]$ and all energies in
units of $k_{\rm B}T$. The free energy densities in Eq.~(\ref{eq:fre_ene})
and Eq.~(\ref{fex}), for example, are in units of $k_{\rm
B}T/[(\pi/6)\bar\sigma^3]$. Also, $\rho_3$ is just the volume fraction
of particles, while $\rho_0$ is a scaled number density, $\rho_0 =
(\pi/6)\bar\sigma^3 n$. With these conventions, the BMCSL contribution
to the excess free energy takes the form
\[
\fex_{\rm BMCSL} = \left(\frac{\rho_2^3}{\rho_3^2} -
\rho_0\right)\ln(1-\rho_3) + 
\frac{3 \rho_1 \rho_2}{1-\rho_3} + \frac{\rho_2^3}{\rho_3(1-\rho_3)^2}\;.
\label{eq:BMCSL}
\]
Given the approximate character of our model free energy, it would not
make sense to try to scale the temperature parameter $t$ precisely to
the temperature in the simulations. Our main aim is to study whether
our $\fex$ can reproduce the qualitative trends observed in the
simulations. While still somewhat crude, our $\fex$ is better suited
to this task than previous versions~\cite{BELLIER00} because it incorporates
polydispersity not only into the attractive contribution, but also
into the hard core reference system.

\subsection{Moment free energy}

Our computational approach for determining the phase behaviour of the
model defined by Eq.~(\ref{fex}) is based on the moment free energy (MFE) method.
We give a brief outline here; details can be found
in~\cite{SOLLICH01,SOLLICH98,WARREN98,SOLLICH02}. The important
feature of our excess free energy is that it is {\em truncatable},
i.e.\ depends only on a finite set of moments $\mi =
\int\!  d\sig\,\wi\rhsig$ defined by weight functions $\wi$; in our case
$\wi=\sig^i$. In coexisting phases, the chemical potentials
$\mu(\sig)$ and pressure $P$ must be equal. The former are, by
differentiation of Eq.~(\ref{eq:fre_ene}),
\be
\musig = \frac{\delta f}{\delta\rhsig} = \ln\rhsig + 
\sum_i \muex_i w_i(\sig), \quad
\muex_i = 
\frac{\partial\fex}{\partial\mi}
\label{musig}
\ee
while the pressure is given by the Gibbs-Duhem relation
\be
P = 
-f + \intsig\musig\rhsig = \rh_0 - \fex + \sum_i \muex_i \mi
\label{P_trunc}
\ee
To the conditions of equality of chemical potentials and pressure we
need to add the requirement of conservation of particle number for
each species $\sig$, which reads
\be
\sum_\alpha x\pa \rh\pa(\sig) = \rhzsig
\label{particle_cons}
\ee
where $\alpha=1,\ldots,p$ labels the phases (compare
equation Eq.~(\ref{eq:vav})). One then finds from
equality of the $\musig$, Eq.~(\ref{musig}), together with particle
conservation Eq.~(\ref{particle_cons}) that the density distributions in
coexisting phases can be written as
\be
\rh\pa(\sig) = \rhzsig \, \frac{\exp\left[\sum_i \lambda_i\pa\wi\right]}
{\sum_\beta x^{(\beta)} \exp\left[\sum_i \lambda_i^{(\beta)}\wi\right]}
\label{rhoalsig_trunc}
\ee
Here the $\lambda_i\pa$ must obey
\be
\lambda_i\pa = -\mu_i^{(\alpha),{\rm ex}} + c_i
\label{lambda_i_exact}
\ee
and the $c_i$ are undetermined constants that do not affect the
density distributions Eq.~(\ref{rhoalsig_trunc}). One can fix them e.g.\
by requiring all the $\lambda_i\pa$ in one of the phases to be zero.
A little reflection then shows that Eq.~(\ref{lambda_i_exact}) together with
$\sum_\alpha x\pa=1$ and the equality of the pressures Eq.~(\ref{P_trunc}) in
all phases give a closed system of nonlinear equations for the
$p(M+1)$ variables $\lambda_i\pa$ and $x\pa$. A solution can thus, in
principle, be found by a standard algorithm such as
Newton-Raphson. Generating an initial point from which such an
algorithm will converge, however, is still a nontrivial problem,
especially when more than two phases coexist and/or many moments $\mi$
are involved. Furthermore, the nonlinear phase equilibrium equations
permit no simple geometrical interpretation or qualitative insight
akin to the construction of phase diagrams from the free energy
surface of a finite mixture.

The moment free energy addresses these two disadvantages. To construct
it, one starts by modifying the free energy decomposition Eq.~(\ref{eq:fre_ene})
to
\be
f = \intsig \rhsig \left[\ln \frac{\rhsig}{\prior}
-1\right] + \fex(\{\mi\})
\label{free_en_decomp}
\ee
In the first (ideal) term, a normalizing factor $R(\sigma)$ has been
included inside the logarithm. This has no effect on the exact
thermodynamics because it contributes only terms linear in $\rhsig$,
but will play a central role below. One can now argue that the most
important moments to treat correctly in the calculation of phase
equilibria are those that actually appear in the excess free energy
$\fex(\{\mi\})$. Accordingly, one imposes particle
conservation Eq.~(\ref{particle_cons}) only for the $\mi$, but allows it to be
violated in other details of the density distribution $\rhsig$ which
do not affect the $\mi$. These ``transverse'' degrees of freedom are
instead chosen to minimize the free energy Eq.~(\ref{free_en_decomp}), and
more precisely its ideal part since the excess contribution is a
constant for fixed values of the $\mi$. This minimization gives
\be
\rhsig=\prior\exp\left[\sum_i \lambda_i\wi\right]
\label{family}
\ee
where the Lagrange multipliers $\lambda_i$ are chosen to give the
desired values of the moments
\be
\mi = \intsig\wi\,\prior\exp\left[\sum_j \lambda_j w_j(\sig)\right]
\label{moments_from_lambda}
\ee
The corresponding minimum value of $f$ as given in Eq.~(\ref{free_en_decomp})
then defines the {\em moment free energy} (MFE)
\be
\fmom(\{\mi\}) = \left(\sum_i \lambda_i\mi - \mze \right) + \fex(\{\mi\})
\label{fmom}
\ee
Since the Lagrange multipliers are (at least implicitly) functions of
the moments, the MFE depends only on the $\mi$. These can now be
viewed as densities of ``quasi-species'' of particles, allowing for
example the calculation of moment chemical
potentials~\cite{SOLLICH01}
\be
\mu_i= \frac{\partial\fmom}{\partial\mi} =
\lambda_i + \frac{\partial\fex}{\partial\mi} = \lambda_i + \muex_i
\label{mom_chem_pot}
\ee
and the corresponding pressure $P = \sum_i \mu_i \mi - \fmom$ which
turns out to be identical to the exact expression Eq.~(\ref{P_trunc}). A
finite-dimen\-sio\-nal phase diagram can thus be constructed from
$\fmom$ according to the usual tangency plane rules, ignoring the
underlying polydisperse nature of the system.  Obviously, though, the
results now depend on $R(\sig)$.  To understand its influence, one
notes that the MFE is simply the free energy of phases in which the
density distributions $\rhsig$ are of the form Eq.~(\ref{family}). To ensure
that the parent phase is contained in the family, one normally chooses
its density distribution as the prior, $R(\sig)=\rhzsig$; the MFE
procedure will then be exactly valid whenever the density
distributions actually arising in the various coexisting phases are
members of the corresponding family
\be
\rhsig=\rhzsig\exp\left[\sum_i \lambda_i\wi\right]
\label{pfamily_precap}
\ee
It is easy to show from Eq.~(\ref{rhoalsig_trunc}) that this condition holds
whenever all but one of a set of coexisting phases are of
infinitesimal volume compared to the majority phase.  Accordingly, the
MFE yields {\em exactly} the onset of phase of coexistence,
conventionally represented via cloud and shadow curves as discussed above.
Similarly, one can show that spinodals and critical points are
found exactly~\cite{SOLLICH01}.

For coexistence involving finite amounts of different phases the MFE
only gives approximate results, since different density distributions
from the family Eq.~(\ref{pfamily_precap}), corresponding to two (or more)
phases arising from the same parent $\rhzsig$, do not in general add
to recover the parent distribution itself. Moreover, from Gibbs' phase
rule, a MFE depending on $M$ moments will not predict more than $M+1$
coexisting phases, while we know that a polydisperse system can in
principle separate into an arbitrary number of phases. Both of these
shortcomings can be overcome by including extra moments within the
MFE; this systematically increases the accuracy of any calculated
phase splits~\cite{SOLLICH01}. By choosing the weight functions of
the extra moments adaptively, the properties of the coexisting phases
can then be predicted with in principle arbitrary
accuracy~\cite{ClaCueSeaSolSpe00,SOLLICH01}. Importantly for us, the
results can in fact be used as initial points from which a solution of
the {\em exact} phase equilibrium problem can be converged
successfully~\cite{SpeSol02,SpeSol03a}. This is the technique that we
use here to obtain results in the coexistence region.

\section{Results}
\label{sec:results}

Here we report the results of applying the simulation and MFE methods
described in sec.~\ref{sec:simmeth} and~\ref{sec:mfe} to obtain the
phase behaviour of size disperse fluids. Our findings are separated
into two parts, those for the sub critical region
(sec.~\ref{sec:subcrit}) and those for the
critical region (secs.~\ref{sec:cp_shifts} and~\ref{sec:top}). With regard to the representation of results, the MFE
calculations are essentially exact for the given model free energy;
symbols indicate the temperatures at which individual calculations were
performed and lines are merely guides to the eye. For the simulation
results, unless otherwise indicated, statistical errors do not exceed
the symbol sizes. Again, lines are guides to the eye.

\subsection{The subcritical region}

\label{sec:subcrit}

Computational complexity limits our investigation of liquid-gas
coexistence to the case of a fluid described by a size distribution of
the Schulz form, Eq.~(\ref{eq:schulz}) having width parameter $z=5$.
This choice of $z$ corresponds to a rather large ($\delta=40.7\%$)
degree of polydispersity. In both the simulations and the MFE
calculation, $f(\sigma)$ was truncated at $\sigma_c=3.0$ and
renormalized appropriately. The simulations were all performed using a
simulation box of volume $V=11390\bar\sigma^3$. In what follows we
quote all simulation quantities in the standard dimensionless LJ units
(e.g.\ $t=k_BT/4\epsilon$). 

\begin{figure}[h]
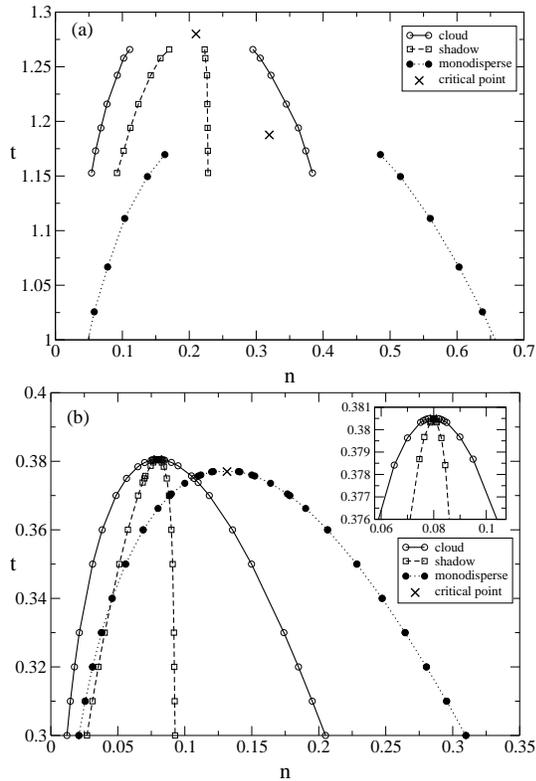

\includegraphics[width=7.0cm,clip=true]{cloud_shadow_dens}
\includegraphics[width=7.0cm,clip=true]{cloud_shad_t_vs_rho}
\caption{ {\bf (a)} Cloud and shadow curves in the $n-T$ plane, as
obtained from the simulations. 
%
%
{\bf (b)} The corresponding prediction of the MFE calculations for a van der
Waals model. Also shown for comparison in each case is a portion of
the coexistence binodal for the monodisperse limit.}
\label{fig:cs_rho}
\end{figure}

We commence by presenting (in fig.~\ref{fig:cs_rho}) a comparison of
the simulation measurements of the cloud and shadow curves in the $n-t$
projection, with those of the MFE calculations.  Apparent in this
representation is a stark separation of the cloud and shadow curves.
Furthermore, the whole phase diagram is considerably shifted with
respect to that of the monodisperse fluid (itself determined previously
in the case of fig.~\ref{fig:cs_rho}(a) in ref.~\cite{WILDING95}).
Specifically, one observes that the critical point occurs at a
considerably higher temperature than in the monodisperse limit. We note
that this particular finding contrasts with that of a previous
theoretical study of a size-disperse van der Waals fluid
\cite{BELLIER00}, which predicts a {\em suppression} of the critical
temperature with respect to the monodisperse limit.

\begin{figure}[h]
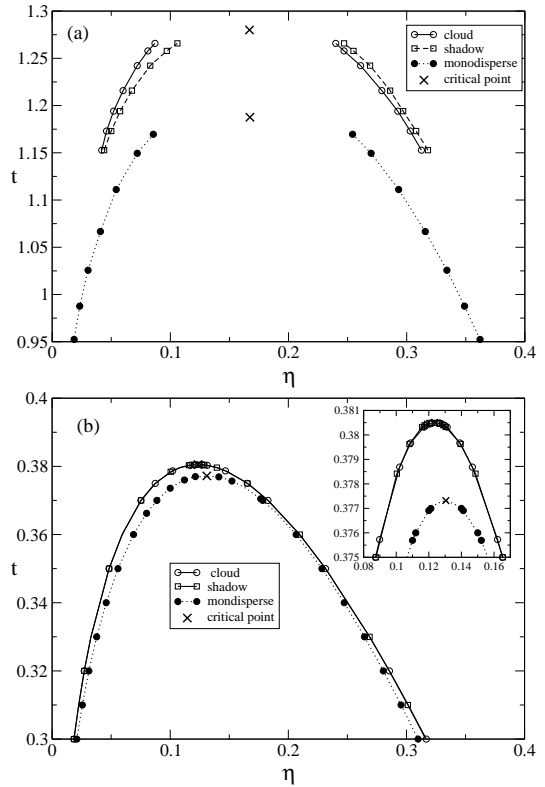

\includegraphics[width=7.0cm,clip=true]{cloud_shadow_eta}
\includegraphics[width=7.0cm,clip=true]{cloud_shad_t_vs_phi}

\caption{{\bf (a)} The simulation, and {\bf (b)} MFE data of
fig.~\protect\ref{fig:cs_rho}, both re-expressed in terms of the volume
fraction $\eta=(\pi/6)\int d\sigma\, \sigma^3\rho(\sigma)$.}
\label{fig:cs_eta}
\end{figure}

\begin{figure}[h]
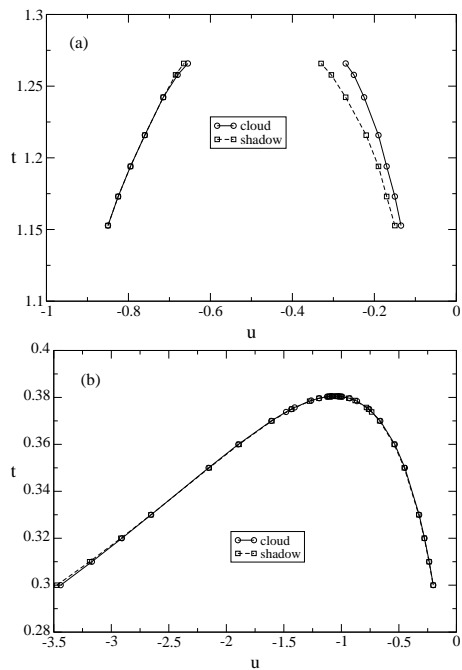

\includegraphics[width=6.0cm,clip=true]{cloud_shadow_ens}
\includegraphics[width=6.0cm,clip=true]{t_vs_e}

\caption{The energy per particle on the cloud and shadow curves
as a function of temperature as obtained from {\bf (a)} the
simulations and {\bf (b)} the MFE calculation. 
%
%
}
\label{fig:cs_ens}
\end{figure}

The order of cloud and shadow curves commonly observed with increasing
density in many polydisperse systems (see e.g.~\cite{PAGON01}) is 
cloud-shadow-cloud-shadow. By contrast, the order shown in
fig.~\ref{fig:cs_rho} is cloud-shadow-shadow-cloud. Interestingly,
however, the order reverts to the standard pattern if one plots the
data in terms of the volume fraction $\eta=(\pi/6)\int d\sigma
\sigma^3\rho(\sigma)$, rather than the overall number density, as shown in
fig.~\ref{fig:cs_eta}. Moreover, one sees that in the $\eta-t$
representation the differences between cloud and shadow phase
properties become much less pronounced. This is particularly evident in the
MFE results, fig.~\ref{fig:cs_eta}(b), for which the cloud and shadow curves
almost coincide, though a close inspection reveals that they are in fact
distinct and occur in the same order as observed in the simulations. Similar
findings pertain to the energy per particle measured
in the cloud and shadow phases (fig.~\ref{fig:cs_ens}), though the
qualitative agreement in the shape of the curves between simulations
and theory is less good here. One reason for this that in our model
free energy the repulsive interactions, which are modelled as hard, do
not contribute to the energy $u$. In the MFE calculations we therefore
determine $u$ as the value of the attractive van der Waals term
from~(Eq.~\ref{fex}). In the simulations, on the other hand, both
attractive and repulsive interactions contribute to the measured value
of $u$.

With regard to the critical point parameters, we note that while the
critical number density of the polydisperse fluid is considerably less
than its value in the monodisperse limit, the simulation estimates of
the critical volume fraction for the mono- and polydisperse fluid agree
to within error. Moreover, in both the simulations and the MFE results,
the critical point is located at or at least extremely close to the
maximum of the cloud and shadow curves. We
see no evidence for portions of the cloud or shadow curves lying above
the critical temperature, which would generically be expected for
polydisperse systems~\cite{SOLLICH02}.
We postpone further discussion of these observations to
sec.~\ref{sec:top}.

\begin{figure}[h]
\includegraphics[width=7.0cm,clip=true]{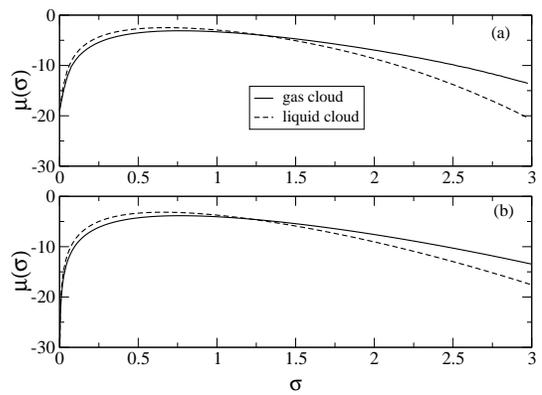}
\caption{The form of the chemical potential distribution $\mu(\sigma)$
at the gas phase and liquid phase cloud points at $t=0.91t_c$. {\bf (a)} Simulation. {\bf (b)}
MFE calculations.}
\label{fig:mu}
\end{figure}

Not all distinctions between cloud and shadow curves can be disguised
by simply recasting the data in terms of $\eta$, rather than $n$. At
temperatures significantly below criticality, we observe considerable
broadening of the coexistence curve in the space of $\mu(\sigma)$. This
is evident in fig.~\ref{fig:mu} which shows the form of $\mu(\sigma)$
at the respective cloud points (marking the boundaries of the two phase
region) for the lowest temperature studied in the simulations,
$t=0.91t_c$. Such 
broadening does not occur in monodisperse systems---coexistence occurs
at a single value of the chemical potential, not a range of values. The
effect is surprisingly large, notwithstanding the high degree of
polydispersity of the parent. Indeed, in simulation terms the
respective cloud points are so far separated in phase space that to
connect them directly (via a route crossing the phase boundary)
required some twenty overlapping simulations---three times the number
required to connect the cloud point to the critical point at this
temperature. 

\begin{figure}[h]
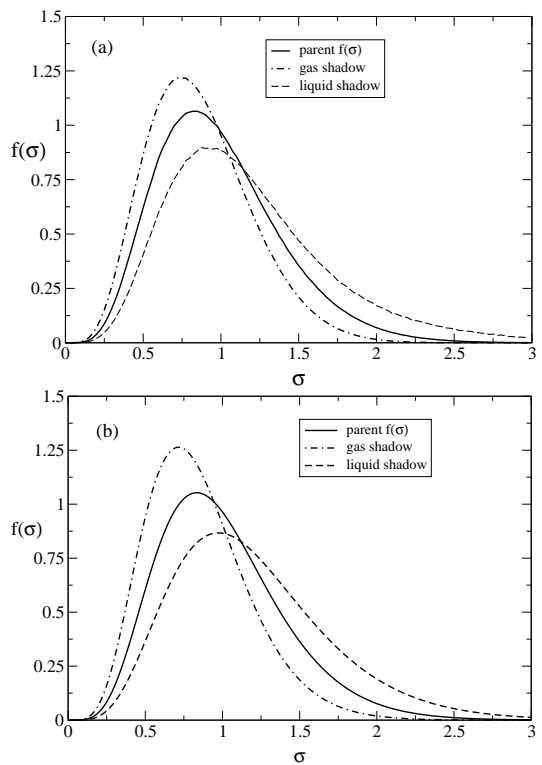

\includegraphics[width=7.0cm,clip=true]{size_dists}
\includegraphics[width=7.0cm,clip=true]{more_size_dists}
\caption{{\bf (a)} Simulation estimates of the normalized daughter
phase density distributions on the shadow curves at $t=0.91t_c$. Also
shown for comparison is the parent distribution. {\bf (b)} The
corresponding results of the MFE calculations.}
\label{fig:s_dists}
\end{figure}

Turning to the form of the density distributions in the shadow phases,
an example of the scale of the difference between these and the parent
is given in fig~\ref{fig:s_dists} for $t=0.91t_c$. The data show that
at the gas phase cloud point, larger particles preferentially occupy
the liquid shadow phase. Conversely at the liquid phase cloud point,
there is a predominance of smaller particles in the gas shadow phase.
Clearly the scale of these fractionation effects is significant. We
note that the principal fractionation effect is a change in the mean
particle diameter. The simulations give for the mean diameter in the
liquid shadow~\cite{terminology} $\langle\sigma\rangle=1.167(3)$, while
in the gas shadow it is $\langle\sigma\rangle=0.863$, both to be
compared to that of the parent, $\langle\sigma\rangle=\bar\sigma=1$.
The changes in the shape of the distribution (as  indicated by
polydispersities of the shadow phases), are rather smaller. For this
temperature ($t=0.91t_c$), we find in the simulations $\delta\approx
43\%$ for the liquid shadow, and $\delta \approx 38.5\%$ for the gas
shadow, to be compared with a parent polydispersity of $\delta=40.7\%$.

\begin{figure}[h]
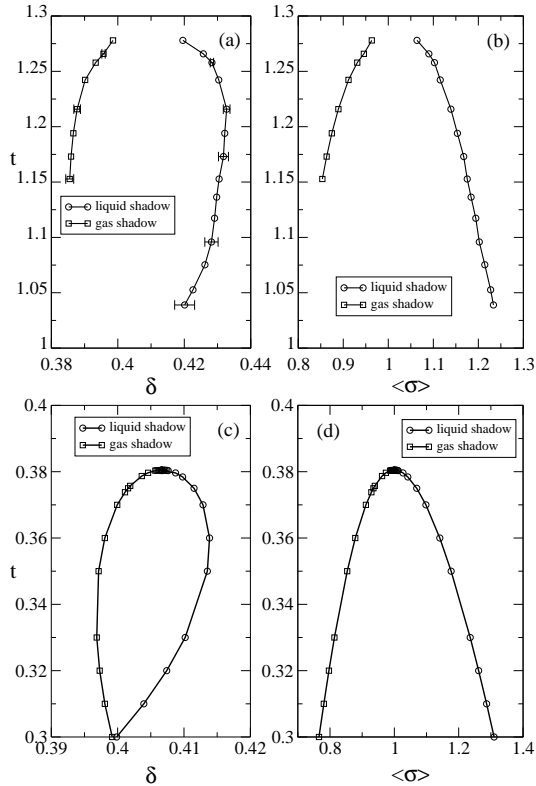

\includegraphics[width=7.0cm,clip=true]{shad_pol_diam}
\includegraphics[width=7.0cm,clip=true]{mor_shad_pol_diam}
\caption{Simulation estimates of the degree of polydispersity
{\bf (a)} and the mean particle diameter {\bf (b)} on the shadow
curves. {\bf (c,d)} The corresponding results of the MFE calculations.}
\label{fig:s_del_diam}
\end{figure}

The temperature dependence of the mean particle size and degree of
polydispersity $\delta$ are shown in fig.~\ref{fig:s_del_diam}. It is
interesting to note that while the value of $\delta$ for the liquid
shadow initially increases strongly with decreasing temperature, on
further reducing temperature it subsequently bends back to lower values of
$\delta$. (Analogous non-monotonic behaviour is observed in the gas shadow.)
Our data (in conjunction with Eq.~(\ref{eq:delta})) show that the origin
of this effect lies in the fact that while
$\langle\sigma\rangle$ increases 
monotonically with decreasing temperature, the standard deviation of
the size distribution in the liquid shadow
phase first increases strongly as $t$ is lowered
from its critical value, but then saturates to a constant value for
temperatures below about $0.9t_c$.

\begin{figure}[h]
\includegraphics[width=9.0cm,clip=true]{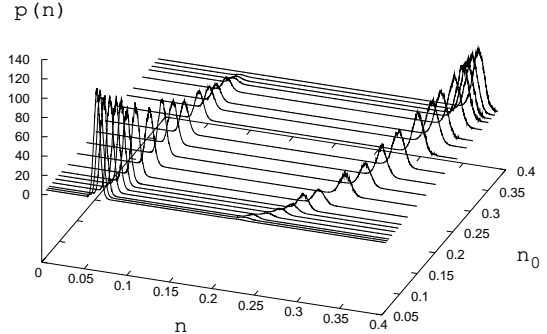}
\caption{Distribution of the fluctuating number density $p(n)$ at a
selection of values of the parent density $n_0$ between the cloud points
at $t=0.91t_c$.}
\label{fig:3Dscan}
\end{figure}

\begin{figure}[h]
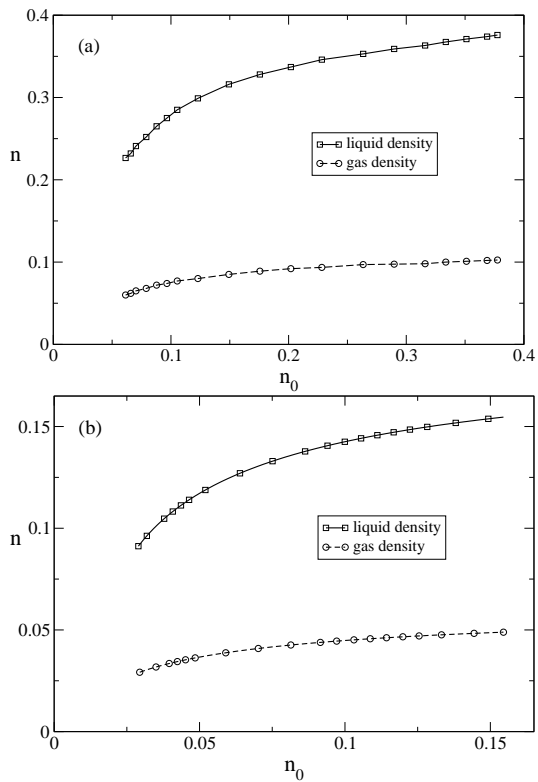

\includegraphics[width=7.0cm,clip=true]{tielinescan}
\includegraphics[width=7.0cm,clip=true]{more_coex_reg.eps}
\caption{Dependence of the gas and liquid densities on the overall
parent number density $n_0$. {\bf (a)} Simulation results extracted from
fig.~\protect\ref{fig:3Dscan}. {\bf (b)} The corresponding results from
the MFE calculations at $t=0.91t_c$.}
\label{fig:lg_dens}
\end{figure}

In order to explore the coexistence region separating the cloud curves,
we have scanned the dilution line at $t=0.91t_c$. The simulations yield
the distribution of the fluctuating number density $p(n)$, the form of
which is shown in fig.~\ref{fig:3Dscan} for a selection of values of
$n_0$ spanning the coexistence region. One observes that in addition to
a transfer of weight between the gas and liquid phase peaks as $n_0$ is
increased from its value at the gas phase cloud point, both the
densities of the gas and liquid peaks shift markedly. This shift of
course reflects the non-coincidence of the cloud and shadow densities
for a given phase. We note further (fig.~\ref{fig:lg_dens}) that the
density shifts are non-linear in $n_0$ and that the overall shift in
the liquid phase density is considerably greater than that for the gas
phase.

\begin{figure}[h]
\includegraphics[width=7.0cm,clip=true]{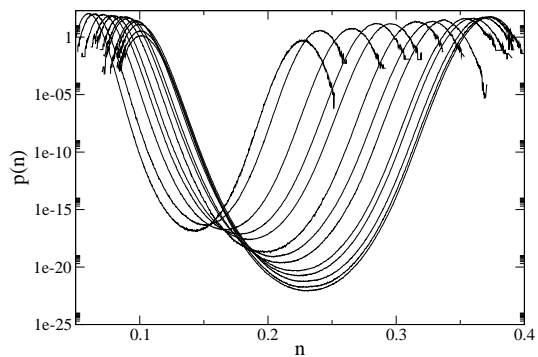}
\caption{The data of fig.~\protect\ref{fig:3Dscan} plotted on a
logarithmic scale 
in order to expose the scale of the free energy barrier separating the
pure phase gas and liquid states in the coexistence region.}
\label{fig:s_tensions}
\end{figure}

Finally in this section, we plot in fig.~\ref{fig:s_tensions} the
measured form of $p(n)$ for state points spanning the coexistence
region, expressed on a logarithmic scale. This representation exposes
the magnitude of the surface tension barrier separating the coexisting
phases, which is proportional to the logarithm of the peak to trough
ratio of the respective distributions \cite{BINDER82}. One observes
that the surface tension of the liquid shadow at the gas phase cloud
point is considerably less than that of the gas shadow at the liquid
phase cloud point. This finding reflects the fact that the liquid
shadow comprises particles that are on average larger than those of
both the parent (cf.\ fig.~\ref{fig:s_dists}) and the gas shadow. It
follows that the number of particles per unit surface area, and hence
the free energy cost of the interface, is smaller between the gas
cloud and the liquid shadow than between the gas shadow and the liquid
cloud.

\subsection{Critical point shifts}
\label{sec:cp_shifts}

Here we enquire how the critical point parameters of the size disperse
fluid depend on the degree of polydispersity $\delta$. To this end we
have employed simulation to investigate four Schulz distribution with
width parameters (cf.\ Eq.~(\ref{eq:schulz})) $z=50,25,10$ and $5$,
corresponding to polydispersities $\delta=14\%, 19.6\%, 30.1\%, 40.7\%$
respectively. For $z=5$ and $z=10$, the size distribution was truncated
at $\sigma_c=3.0$, while for $z=25$ and $z=50$ a cutoff $\sigma_c=2.0$
was used. 

Standard finite-size scaling (FSS) methods were used to estimate the
location of the critical point in each case studied. The FSS methodology
exploits the critical point scale invariance of the fluctuation
spectrum of quantities such as the number density, as expressed through
the form of $p(n)$ \cite{WILDING95}. For the $z=50$ and $z=25$
distributions, three system volumes were studied, namely
$V=3375\bar{\sigma}^3, 8000\bar{\sigma}^3, 15625\bar{\sigma}^3$ . For
the $z=10$ and $z=5$ distributions, somewhat larger simulation cells
were required, the values chosen being $V=11390\bar\sigma^3,
27000\bar\sigma^3$, and $V=52734\bar\sigma^3$. We note that the largest
of these systems contained over $10^4$ particles at criticality. 

\begin{figure}[h]
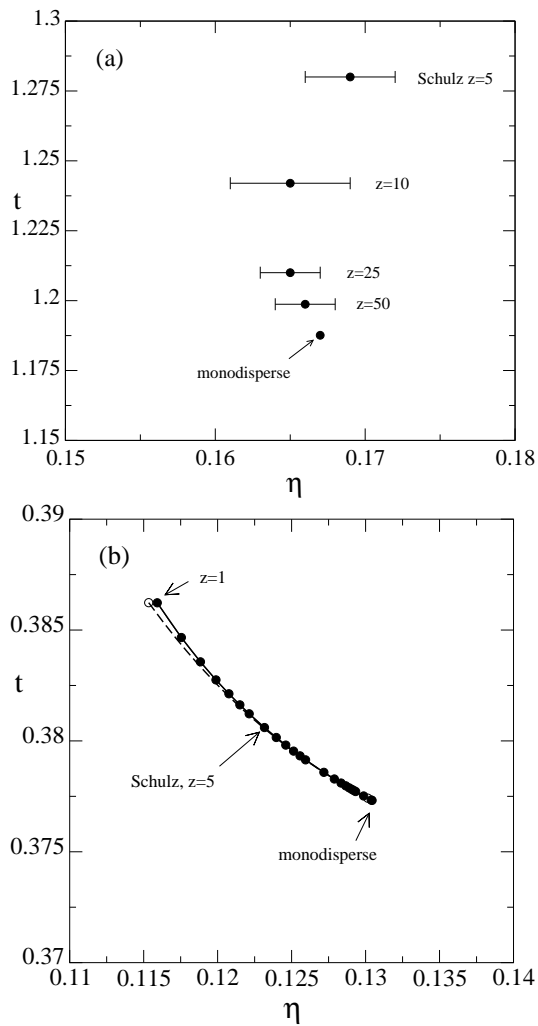

\includegraphics[width=7.0cm,clip=true]{critparams}
\includegraphics[width=7.0cm,clip=true]{BMCSL_critical_points}
\caption{{\bf (a)} Simulation estimates of the critical temperature and
volume fraction for Schulz parent distributions of a variety of widths
(see text for details). {\bf (b)} Corresponding predictions of the MFE
calculations for Schulz distributions in the range $z=1$ to $z=\infty$
(monodisperse limit). The dashed line and empty circle indicate the
location of the spinodal maximum; see sec.~\protect\ref{sec:top}.}
\label{fig:crit_pars}
\end{figure}

In order to compare with the simulation results, MFE calculations were
performed to obtain the critical point for width parameters in the
range from $z=1$ to the monodisperse limit $z=\infty$. The resulting
comparison is shown in fig.~\ref{fig:crit_pars}. We note that while the
simulations were able to locate the critical temperature $t_c$ rather
accurately (the error bar is smaller than the symbol size), the
estimates of the critical volume fraction $\eta_c$ carry a rather large
estimated error  because of the difficulty in obtaining good statistics
for large systems and high degree of polydispersity. Thus, although we
find a strong increase in $t_c$ with increasing $\delta$, no
clear trend is observed for $\eta_c$. The MFE results confirm the
increase of $t_c$ with increasing $\delta$. However, the fractional
change is substantially less than that observed in the simulations. As
regards the critical volume fraction, the MFE method predicts a
significant ($5\%$) decrease in the value of $\eta_c$ over the range of
$\delta$ studied in the simulations. The scale of this decrease is,
however, larger than any that might be considered consistent with the
simulation error bars.

\subsection{Location of the critical point}
\label{sec:top}

An important finding of our investigation of critical point shifts is
that, for each $z$ studied, the critical point occurs at, or very
close to, the maximum of the cloud and shadow curves. Correspondingly,
there is no numerical evidence of distinct cloud and shadow points at
or above $t_c$, either in the simulations or the MFE
calculations. This observation is consistent with earlier results
obtained from a simpler model free energy~\cite{BELLIER00}, but
clearly calls for an explanation: in a polydisperse system the
critical point must be found at an intersection of the cloud and
shadow curves, and this intersection will in general be located {\em
below} the maxima of the two curves~\cite{SOLLICH02,RASCON03}.

One possible way to investigate when the critical point is near the
top of the cloud (and hence also the shadow) curve is to find
conditions under which the slope of the cloud curve there is
zero. This can be done for simple excess free energies depending only
on a single density moment, within the framework of a Landau
expansion~\cite{RASCON03}. However, such an approach becomes
excessively unwieldy when several density moments are involved. We
have therefore addressed the question in a more indirect way. On
general grounds the spinodal curve, where a given polydisperse phase
becomes locally unstable, must lie inside the cloud curve and touch it
at the critical point~\cite{SOLLICH02}. This implies that the critical
point is at the maximum of the cloud curve exactly when it coincides
with the maximum of the spinodal (as happens in monodisperse
systems). We can therefore estimate the shift of the critical point
away from the cloud curve maximum by comparing its location with that
of the spinodal maximum: as long as these two points are close
together, the critical point should also be close to the cloud curve
maximum. 

The advantage of this approach is that both the spinodal curve and the
critical point can be calculated relatively easily within the MFE
framework, from local properties of the excess free energy
surface. Fig.~\ref{fig:crit_pars} shows that the resulting locations
of the critical point and spinodal maximum are, though not identical,
extremely close for the whole range of width parameters $z$ studied
above. Even for the most polydisperse system ($z=1$,
$\delta=1/\sqrt{2}\approx 70\%$) the temperature coordinates of the two
points are indistinguishable on the scale of the graph, confirming our
earlier observation that the critical point is very near the cloud
curve maximum.

To understand under which circumstances such rather unusual behaviour
can be expected, we have derived general expressions for the location
of the critical point and spinodal maximum in a polydisperse system with a
smooth (van der Waals-type) excess free energy. Details will be given
in a forthcoming publication~\cite{SOLLICH04}. Within an expansion
valid for near-monodisperse systems, i.e.\ small polydispersity
$\delta$, the leading shifts in the critical point and spinodal
maximum relative to the monodisperse limit $\delta\to 0$ turn out to
be $O(\delta^2)$.  The density shifts are in general {\em different}
as expected on general grounds for a polydisperse system; see
above. (The temperature shifts, which are essentially quadratic in the
density shifts because we are expanding around the maximum of the
monodisperse spinodal curve, are always identical to $O(\delta^2)$.)
However, it turns out that for purely size-polydisperse systems, where
the excess free energy is unchanged if we scale all particle sizes and
interparticle separations by the same factor, the lowest-order density
shifts of critical point and spinodal maximum {\em exactly}
coincide. Likewise, one can show using the techniques
of~\cite{EVANS98,EVANS01} that size polydispersity causes the lowest-order
shifts in the entire cloud and shadow curves to be identical, once
they are plotted in terms of volume fraction rather than density. This
is consistent with our observation of the near-coincidence of these
curves in fig.~\ref{fig:cs_eta}. 

Thus the perturbative expansion shows that the effects which we observe
numerically are due to the fact that our LJ system has purely
size-polydisperse interactions. This rationalizes why earlier
calculations~\cite{BELLIER00} also found the critical point very near
the top of the cloud curve, even though a much simpler free energy
model was used: any excess free energy which respects the scale
invariance of a size-polydisperse system will exhibit such behaviour.

\begin{figure}[h]
\includegraphics[width=7.0cm,clip=true]{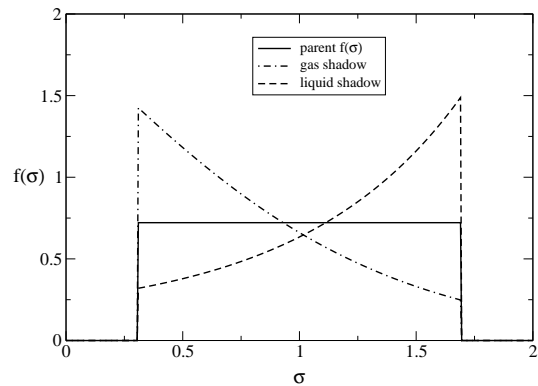}
\caption{Normalized size distributions
in the daughter phases on the shadow curves at $t\approx0.9t_c$, for a
top hat parent distribution with polydispersity $\delta=40\%$.
\label{fig:distr_top_hat}
}
\end{figure}

\begin{figure}[h]
\includegraphics[width=8.0cm,clip=true]{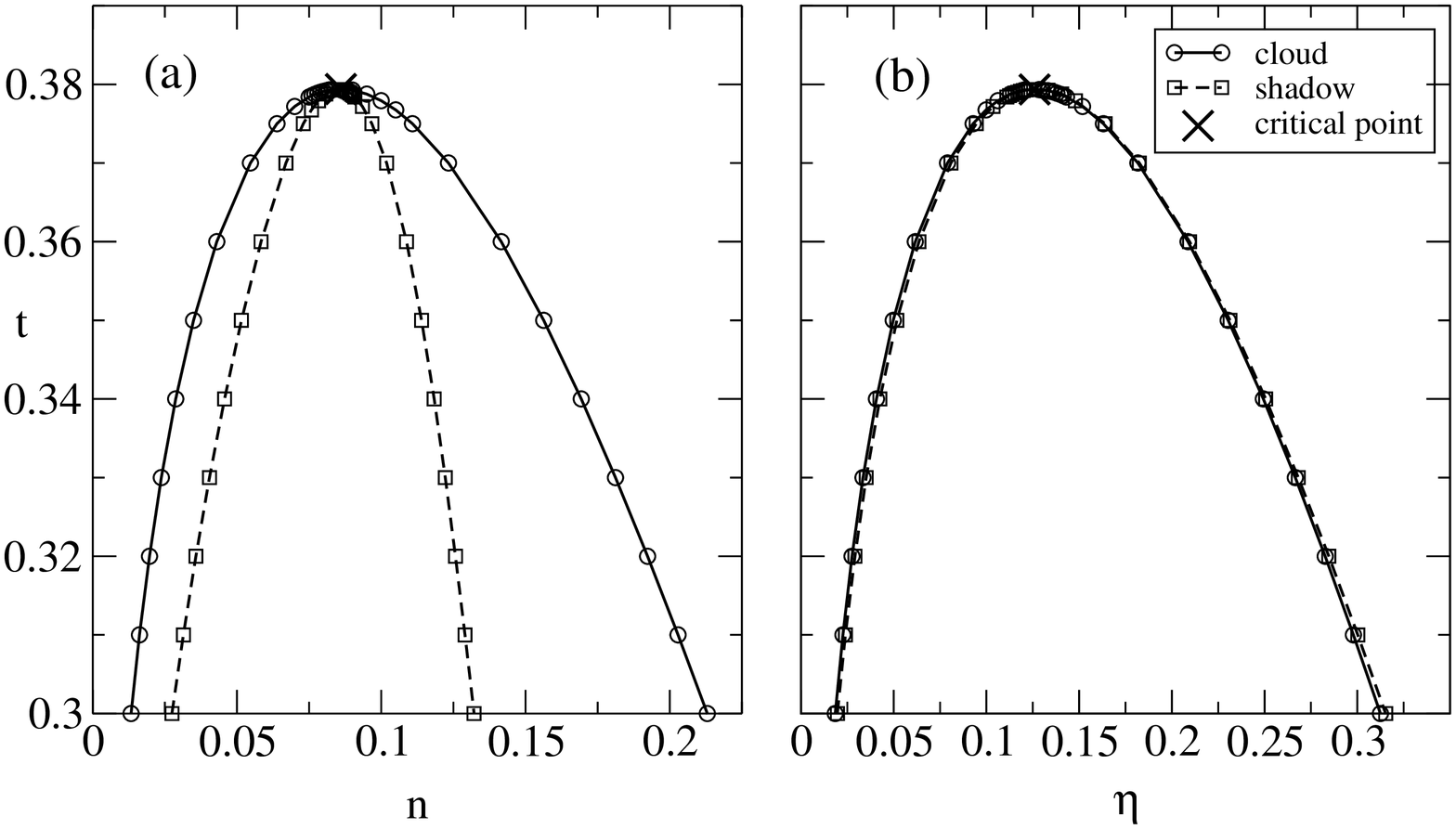}
\includegraphics[width=8.0cm,clip=true]{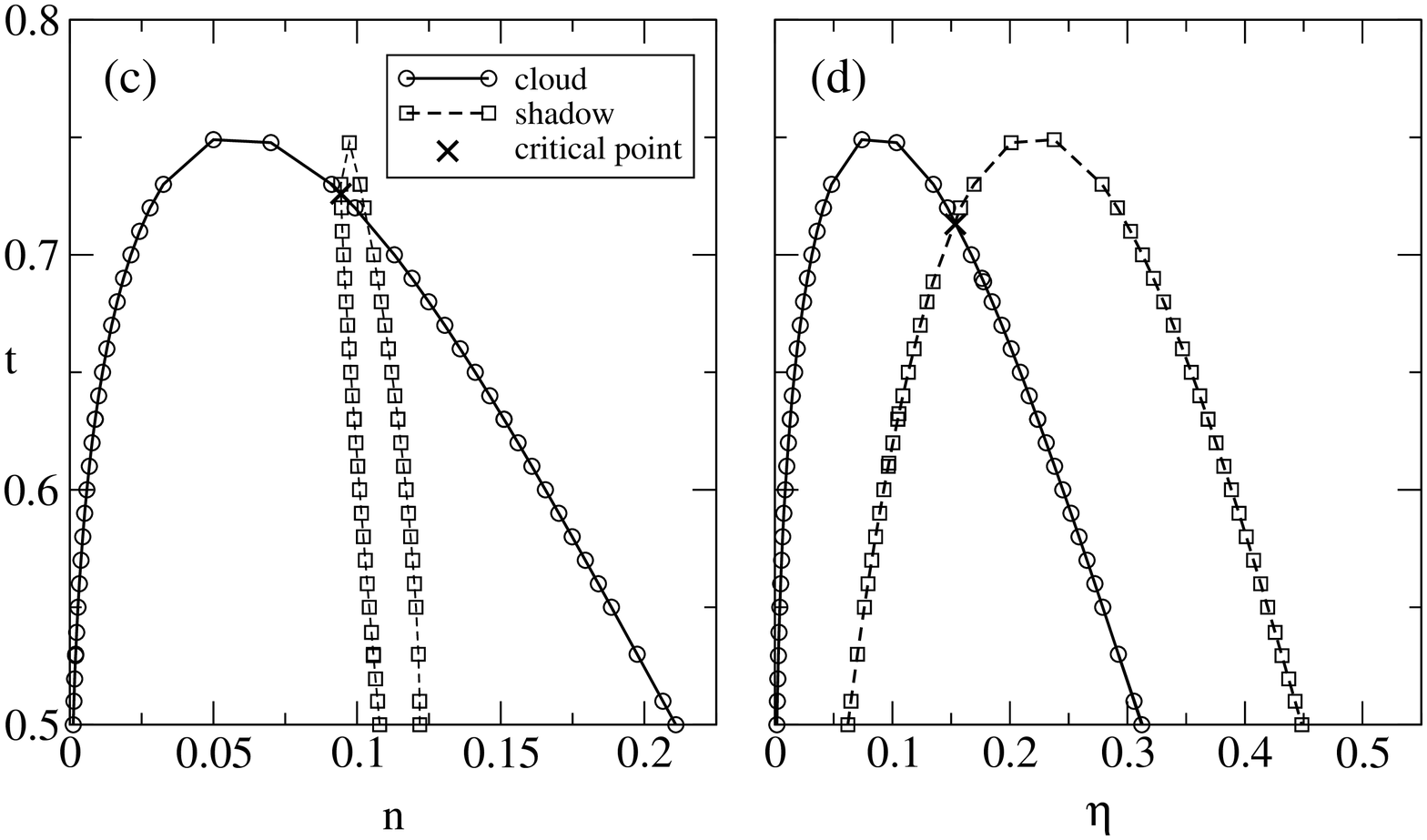}
\caption{{\bf (a,b)} Cloud and shadow curves in the $n-t$ and $\eta-t$
planes, for a parent with a top-hat size distribution, as obtained by
MFE calculations. {\bf (c,d)} The analogous results for a system with
added amplitude polydispersity; the critical point is now clearly
below the maximum of cloud and shadow curves.
\label{fig:cloud_shadow_top_hat}
}
\end{figure}
We initially thought that the fact that our critical points were at
the top of the cloud curves was related to the size distributions in
the shadow phases being essentially just scaled versions of the cloud
(parent) size distributions; see fig.~\ref{fig:s_dists}. Indeed, one
can show that for any size-polydisperse excess free energy
which depends only on the density moments $\rho_0$ and
$\rho_1$, this scaling holds {\em exactly} when the parent phase has a
Schulz size distribution~\cite{SOLLICH04}. In such a scenario the
cloud and shadow curves also coincide exactly in the $\eta-t$
representation, and the critical point is exactly at their common
maximum~\cite{SOLLICH04}. However, the same features can occur, to a
very good approximation, even when there is no scaling link between
cloud and shadow size distributions. We demonstrate this by
considering theoretically a parent phase with a top hat size
distribution, with the polydispersity $\delta=40\%$ taken to be
the same as our main Schulz distribution example.
Fig.~\ref{fig:distr_top_hat} shows exemplary size distributions in the
shadow phases: these are now clearly different in shape from each
other and from the parent. Nevertheless
(fig.~\ref{fig:cloud_shadow_top_hat}(a,b)) the $\eta-T$ cloud and shadow
curves are almost identical and the critical point is extremely close
to their maximum. 

Finally, the above theoretical arguments suggest that, once we move to
a case where polydispersity affects not only the size but also the
amplitude of the interaction potentials, the behaviour expected for a
generic polydisperse system should be recovered. To confirm this, we
repeated the MFE calculations for interaction potentials $U_{ij}$,
Eq.~(\ref{Uij}), scaling as the product $\sigma_i \sigma_j$ of the
sizes of the particles involved. This translates into an additional
factor $\sigma\sigma'$ in the attractive term in the model free
energy~(Eq~.\ref{fex_first}). The results are shown in
fig.~\ref{fig:cloud_shadow_top_hat}(c,d) for the top hat parent: as
expected, the critical point is now clearly below the cloud curve
maximum, and the $\eta-T$ representations of the cloud and shadow
curves no longer coincide. Preliminary simulation results for a LJ
model with combined size and amplitude polydispersity confirm this
scenario \cite{WILDING04_unpub}.

\section{Discussion and outlook}
\label{sec:disc}

In summary, we have used specialized MC simulation and the moment free
energy (MFE) method to study the liquid-gas phase behaviour of a
size-disperse LJ fluid having a degree of polydispersity
$\delta=40.7\%$.  Cloud and shadow curves have been traced and
fractionation effects quantified. Surprisingly good qualitative
agreement is found in almost all aspects of the results between the
simulations and the MFE calculations for a van der Waals model.

In a related study, we have also obtained the dependence of the
critical point parameters on the degree of polydispersity. The
simulation show a strong rise in the critical temperature with
increasing $\delta$, but no clear trend in the critical volume fraction
$\eta_c$. The increase in $t_c$ is confirmed by the MFE calculation,
albeit with a weaker magnitude. However, in contrast to the simulation,
the MFE predicts quite a strong decrease of $\eta_c$ with increasing
$\delta$.

Some of our findings for the sub-critical coexistence region seemed,
at first sight, rather surprising: the cloud and shadow curves nearly
collapsed onto a common curve in the $\eta-t$ representation, with the
critical point found so close to their maximum as to be
indistinguishable. We outlined a theoretical argument which shows that
such behaviour is generically to be expected for systems with pure
size-polydispersity, but not otherwise. This was confirmed by, on the
one hand, considering a different parent size distribution, with a top
hat shape: similar behaviour was observed. The addition of amplitude
polydispersity, on the other hand, completely changed the picture.
We demonstrated finally that the close similarity in the shapes of the
size distributions of cloud and shadow phases is rather peculiar to
parents with a Schulz size distribution, and not directly linked to
the occurrence of the other effects of size polydispersity which we
observe.

Given the rather crude nature of the van der Waals free energy Eq.~(\ref{fex}),
used for the MFE calculation, the level of qualitative agreement with
the simulations in the sub-critical region is gratifying. We note,
however, that previous work treating polydispersity within a van der Waals
framework \cite{BELLIER00} did not produce a similar level of
agreement. We tentatively ascribe this difference to the inclusion in
the present work of an explicitly polydisperse hard-core reference free
energy having the well known BMCSL form \cite{BOUBLIK,MANSOORI}. Use of
this reference free energy improves the description of packing effects.
These are expected to be particularly important for the correct
description of liquid structure in polydisperse fluid because of the
possibility that small particles can fit into the gaps between large
ones. We speculate that it is this ability to pack more effectively
(and the resulting lowering of the configurational energy) which is
responsible for the observed increase in $t_c$ with increasing
polydispersity, a finding which contrasts with the predicted {\em
decrease} in ref.~\cite{BELLIER00}.

One surprise arising from the simulation results is the observed
magnitude of coexistence curve broadening. In simulations of a
monodisperse system at coexistence, the system fluctuates freely
between both phases (assuming the provision of an appropriate
multicanonical weight function) \cite{WILDING95}. The form of $p(n)$
exhibits two peaks, and these are visible over a certain range of $\mu$
due to finite-size smearing of the transition. By contrast the range of
$\mu(\sigma)$ (fig.~\ref{fig:mu}) over which gas and liquid phase peaks
are visible (cf. fig.~\ref{fig:3Dscan}) is $1$-$2$ orders of magnitude
greater than the finite-size smearing one would expect in a
monodisperse system having a similar number of particles. Consequently
a large number of separate, but overlapping (in configuration space)
simulations are required to bridge the coexistence region from pure gas
to pure liquid.

With regard to the general simulation issues raised in this study, we
have shown that recently developed techniques (dilution line tracking
\cite{WILDING02} bootstrapped by the non-equilibrium potential
refinement method \cite{WILDING03}) can be successfully employed to map
the phase behaviour of fluids exhibiting fixed polydispersity.  This is
achieved within what is arguable the most flexible framework for the
study of fluid phase equilibria, namely the grand canonical ensemble. 
Notwithstanding these advances, however, it should be stressed that the
simulations themselves are not yet routine: the work reported here
consumed well over $10^4$ hours of CPU time on a $2$ GHz PC processor.
Such an investment of effort exceeds that necessary for obtaining the
phase behaviour of a monodisperse fluid by a factor of $1$-$2$ orders
of magnitude. The source of the computational complexity is twofold.
Firstly, one needs to utilize system sizes whose linear extent exceeds
a given factor of the largest possible particle diameter. Even for
moderate polydispersities, this can increase the necessary volume by an
order of magnitude compared to the monodisperse case. Secondly, in
order to relax the system between sampled configurations, one must
decorrelate not only the overall density, but also the distribution of
particle sizes $\rho(\sigma)$. The bottleneck for the latter process is
the largest particles, for which the probability of particle insertions
and deletion is small, and which can thus only be altered by the
resizing moves whose effect is generally more incremental in magnitude.
Clearly there is scope for further algorithmic improvements in this
regard.

As an extension of the present study, work is under way to simulate the
phase diagram of a Lennard-Jones fluid in which the well depth in the
LJ interaction potential depends on the sizes of the interacting
particles. Preliminary results \cite{WILDING04_unpub} confirm the
prediction of the MFE calculations described in sec.~\ref{sec:top},
i.e. that the cloud and shadow curves adopt the standard order and that
the critical point lies well below the extremum of the cloud and shadow
curves.

\acknowledgments

The authors acknowledge support of the EPSRC, grant numbers
GR/S59208/01 and GR/R52121/01.

\end{document}